\documentstyle[12pt,draft,epsf]{article}

 \newcommand {\bi} {\bibitem}
 \newcommand {\be} {\begin{equation}}
 \newcommand {\bd} {\begin{math}}
\newcommand {\bea} {\begin{eqnarray} \nonumber }
\newcommand {\ee} {\end{equation}}
\newcommand {\ed} {\end{math} }
\newcommand {\eea} {\end{eqnarray}}
 \newcommand {\eps} {\epsilon}
 \newcommand {\si} {\sigma}
\newcommand {\de} {\delta}

\newcommand {\ga} {\gamma}
\newcommand {\la} {\lambda}

 \newcommand {\al} {\alpha}

\newcommand {\ba} {\overline}

\newcommand{\nn}{\nonumber\\}

\def \form#1 {eq. (\ref{#1}) }
\def \parziale#1#2  {{\partial {#1} \over \partial {#2}}}
\def\(({\left(}
 \def\)){\right)}
\def\[[{\left[}
\def\]]{\right]}

\def\bW1{{\bf  W_1}}
\def\bh1{{\bf  h_1}}

\begin{document}

\title{On the replica approach to glasses}
\author{  Giorgio Parisi\\
Dipartimento di Fisica, Universit\`a {\em La  Sapienza},\\ 
INFN Sezione di Roma I \\
Piazzale Aldo Moro, Rome 00185}
\maketitle

\begin{abstract}
Here we review the approach to glassy systems based on the replica method and we 
introduce the main ingredients of replica symmetry breaking.  We explain why the replica method has 
been successful in spin glass and why it should be successful for real glasses.
\end{abstract}

\section {Introduction}

The idea that glasses and spin glasses have something in common (beyond the name) is often stated in 
the literature.  We can hope that the progresses that have been done in these last 20 years in the 
study spin glasses can be exported to glasses \cite{mpv,parisibook2,MPR}.  Many of these progresses 
have been done using the replica method with the assumption of spontaneous symmetry breaking.

The aim of this note is to describe the foundations of the replica method and to show why glasses 
are well inside its scope.  It is organized as follows: in section II we review the replica method 
and we introduce the formalism of replica symmetry breaking.  In section III we recall some mean 
field results that have been obtained for infinite range generalized spin glasses.  In section IV we 
present some considerations on real glasses and in section V we explain why the replica method 
should work for real glasses.  Finally in the appendix we present a first computation of the 
properties of a soft sphere glass using the replica method.

\section{The replica method}
\subsection{Why replicas}
If we deal with systems which order in a simple way at low temperature (i.e.  crystal forming 
materials, ferromagnets), the life is simple: we have to find the ground state and study the 
fluctuations around the ground state.

The situation is much more difficult when the for one reason or another (intrinsic randomness, 
chaotic dependance on the size of the system) the ground state is not known.  In this case it is 
convenient to introduce replicas of the system.  For example in spin glasses, where the fundamental 
variables are Ising spins which are defined on the points of the lattice, one introduces $n$ 
replicas of the same system \cite{mpv,parisibook2}, with Hamiltonian
\be
H=\sum_{a=1,n} H[\si^a].
\ee

Let us consider the case where at low temperature the system develops a spontaneous magnetisation, 
which change sign from site to site with zero average, i.e.
\be
<\si_a(i)>=m(i).
\ee
In this case the system average of the magnetisation, i.e. 
\be
m \equiv {\sum_{i=1,N}<\si_a(i)> \over N},
\ee
is not informative. On the contrary if we look to a two replicas quantity
(for $a\ne b$), we find the rather interesting results:
\be
q_{a,b}\equiv {\sum_{i=1,N}<\si_a(i)\si_b(i)> \over N} = {\sum_{i=1,N}m(i)^2 \over N}.
\ee
In the nutshell we use one replica to probe the properties the other replica.  Correlations 
functions among different replicas (averaged over the system) bring information to us on the 
properties of the ground state of the system although they are not sufficient to determine the 
ground state.

Using this formalism we accomplish two goals:
\begin{itemize}
\item If we measure the correlations among replica and not the ground state 
directly, we have quantities which should be more stable when we perturb the 
system.
\item If we succeed to write down theoretical relations for the correlations 
among different replicas, we can use them to solve the model without having to 
compute the ground state.
\end{itemize}

\subsection{The breaking of replica symmetry}
 
Things are more complicated if we have more than one ground state.  For example for Ising spins in 
absence of magnetic field there are always two possible values of the magnetisation, which are 
obtained by changing the sign to all the spins.

Usually we cure this problem by adding a small magnetic field, but this does not always solve the 
problem, e.g.  it does not solve the problem in the case of an antiferromagnet.  In this case one 
finds that
\be
<\si_a(i)\si_b(i)>=0,
\ee
because the four (two for the replica $a$ and two for the replica $b$) states of give an exactly equal in 
modulus and cancelling contribution.

The solution is simple: for example in the case $n=2$ we consider the following Hamiltonian 
\cite{CAPAPASO90,KPVI,KLEIN,FP} :
\be
H=H(\si_{1})+H(\si_{2})- \eps \sum_{i}\si_{1}(i)\si_{2}(i),
\ee
and we study its properties for small $\eps$.

Now the Hamiltonian is invariant under the change of {\em both} the first and the second replica 
together.  If there are two states, having magnetisation $\pm m(i)$ one finds
\bea
\lim_{\eps \to 0^{+}}q(\eps)=-\lim_{\eps \to 0^{-}}q(\eps)={\sum_{i=1,N}m^{2}(i) \over N},\\
q(\eps)\equiv{\sum_{i=1,N}<\si_1(i)\si_2(i)>_{\eps} \over N}.
\eea

By adding a term which couples different replicas and observing if the correlations among replicas are 
continuous or discontinuous when this parameter change sign, we can monitor the existence of one of 
more equilibrium states.  Just because the equivalence among replicas is destroyed by an 
infinitesimal field where more than one equilibrium state is present, we say that in this situation 
replica symmetry is spontaneously broken \cite{mpv,parisibook2}.

It is evident that if we should consider only the case in which the equilibrium states differ by a 
simple symmetry transformation, the replica formalism would be correct, but may be not so useful.  
The real power of the replica formalism shows up when there are many equilibrium states of different 
free energy.  In one of the most simple non trivial cases which can be studied by using this 
formalism, there are many equilibrium states and the probability of finding a state with free energy 
$F=\Delta F + F_{0}$ ($F_{0}$ is the free energy of the lowest lying state) increases as
\be
P(\Delta F) \propto \exp (\beta m \Delta F), \label{FREE}
\ee
where the parameter $m$ characterize the way in which the replica symmetry is broken 
\cite{mpv,parisibook2}.
This case is usually  called one step breaking.

There is an algebraic formulation of the replica method in which one encodes the information coming 
from the structure of the states in correlations among replicas.  The $n$ replicas are divided in 
groups in such a way that correlations among replicas belonging in the same group are the same.  For 
example in the previous case of one step replica symmetry breaking, one divides the $n$ replicas in 
$n/m$ groups of $m$ replicas each.  This construction may look rather artificial, but it turns out 
to be an extremely efficient way of describing in a compact way rather complex situations.  Moreover 
this formulation is well suited for doing computations \cite{mpv,parisibook2}.  It may take some work 
to do the decoding and to find out which are the physical properties of the state distribution from 
the assumed for of replica symmetry breaking; however this decoding is always possible and the 
structure of the probability distribution of the states mirrors the structure of the algebraic 
properties of the replicas.
\section{Mean field results}
In these recent years there have been many progresses on the understanding of the behaviour of 
glassy systems in the mean field approximation.  The main results have been obtained for the 
following cases:
\begin{itemize}
\item Models with random quenched disorder have been well understood also from
the dynamical point of view.  

\item Models with random quenched disorder, which display a glassy transition quite similar to
that of the previous models. Some of the results obtained for systems with random quenched disorder 
have been extended also to these systems.
\end{itemize}
Let us see what happened in more details.

\subsection{Systems with quenched disorder}

Generally speaking when the interaction range is infinite, the thermodynamical properties at 
equilibrium of a spin model can be computed analytically.  When the system is disordered, we must 
use the replica method \cite{mpv,parisibook2}.  A typical example of a model which can be solved 
with the replica method is a spin model with $p$ spin interaction.  The Hamiltonian we consider 
depends on some control variables $J$, which have a Gaussian distribution and play the same role of 
the random energies of the REM and by the spin variable $\si$.  For $p=1,2,3$ the Hamiltonians are 
respectively
\begin{eqnarray}
H^1_J(\si)= \sum_{i=1,N} J_i \si_i;\ \ \ \ \ \ 
H^2_J(\si)= \sum_{i,k=1,N}' J_{i,k} \si_i \si_k;
\ \ \ \  \ \ H^3_J(\si)= \sum_{i,k,l=1,N}' J_{i,k,l} \si_i \si_k \si_l\ ,
\end{eqnarray}
where the primed sum indicates that all the indices are different \cite{GROMEZ,GARDNER}.  The $N$ 
variables $J$ must have a variance of $O(N^{(1-p)/2})$ in order to have a non trivial 
thermodynamical limit.  The variables $\si$ are usual Ising spins, which take the values $\pm 1$.  
From now on we will consider only the case $p>2$.

In the replica approaches one assumes that at low temperatures the phase space 
breaks into many valleys, (i.e.  regions separated by high barriers in free 
energy).  One also introduces the overlap among valley as
\be
q(\al,\ga)\equiv {\sum_{i=1,N} \si^\al_i \si^\ga_i \over 2 N},
\ee
where $\si^\al$ and $\si^\ga$ are two generic configurations in the valley 
$\al$ and $\ga$ respectively.

In the simplest version of this method \cite{GROMEZ,GARDNER} one introduces the typical overlap of 
two configurations inside the same valley (sometimes denoted by $q_{EA}$).  Something must be said 
about the distribution of the free energies of the valleys.  Only those which have minimum free 
energy are relevant for the thermodynamics.  One finds that these valleys have zero overlap and have 
the following distribution of the {\em total} free energy $F$:

\be P(F)\propto \exp(\beta m (F-F_0)).  \label{DEFM}\ee

The two parameters, $q$ and $m$ give complementary information:
\begin{itemize}
\item The parameter $q$ tell us how much two generic configurations in the same valley are similar
one to the other; with this normalization $q$ should go to 1 to zero temperature.
\item 
The parameter $m$ tell us how much the different valley differs in free energy.  If we assume that the 
free energy differences remain finite when the temperature goes to zero, $m$ is proportional to the 
temperature at small temperature.
\end{itemize}

The two parameters $q$ and $m$ are enough to characterize the behaviour the system. Indeed the 
average value of the free energy density ($f$) can be written in a self consistent way as function of 
$m$ and $q$ ($f(q,m)$) and the value of these two parameters can be found as the solution of the 
stationarity equations:

\be \parziale{f}{m} 
=\parziale{f}{q} =0.  \ee

The quantity $q$ is of order $1-\exp(-A\beta p)$ for large $p$, while the parameter $m$ is 1 at the 
critical temperature, and has a nearly linear behaviour al low temperature.  

The thermodynamical properties of the model are the same as is the Random Energy Model 
\cite{REM} (indeed we recover the REM when $p\to\infty$ \cite{GROMEZ} ): there is a transition
at
$T_{c}$
with a
discontinuity in the specific heat, with no divergent susceptibilities.

A very interesting finding is that if we consider the infinite model and we cool it starting at high 
temperature, there is a dynamical transition at a temperature $T_{D}>T_{c}$ \cite{KT1}-\cite{BCKM} .  
At temperatures less than $T_{D}$ the system is trapped in a metastable state with energy greater 
than the equilibrium energy.  The existence of these {\em infinite} mean life metastable states is 
one of the most interesting results for these models. The correlation time (not the equilibrium 
susceptibilities) diverges at $T_{D}$ and the mode-mode coupling become exact in this region.

\subsection{Systems without quenched disorder}
We could ask how much of the previous results can be carried to models without quenched results.  It 
has been found in the framework of the mean field theory (i.e.  when the range of the interaction is 
infinite), that there a the partial equivalence of Hamiltonians with quenched and random disorder.  
More precisely it often possible to find Hamiltonians which have the same properties (at least in a 
region of the phase space) of the Hamiltonian without disorder \cite{MPR} - \cite{CKPR} .  An 
example of this intriguing phenomenon is the following.

The configuration space of our model is given by $N$ Ising spin variables \cite{MPR}. 
We have a line of line of length $N$ and on each site there is an Ising variable $\si(i)$.
The Hamiltonian is a four spin antiferromagnetic interaction
\be
H=N^{-1}\sum_{i,k,j,l}\si(i)\si(k)\si(j)\si(l),
\ee
where the sum is restricted to the sites that satisfy the condition $i+k\equiv_{N}j+l$ (where 
$\equiv_{N}$ stand for congruent modulus $N$).  This Hamiltonian has been proposed in the study of low 
autocorrelation sequences.  With some work it can be rewritten as:
\begin{eqnarray}
H=\sum_{i=1,N}|B_i|^2-1|^2, \\
\mbox{where} \ \ \ B_i=\sum_{k=1,N} R_{i,k} \si_k.
\end{eqnarray}
Here $R$ is an unitary matrix, i.e.
\be
\sum_{k=1,N}R_{i,k}\ba{R_{k,m}}=\de_{i,m}.
\ee
which in this particular case is given by
\be
R(k,m) ={\exp (2 \pi i \ k m) \over N^{1/2}}\label{FOU}
\ee

 We could consider two different cases \cite{MPR} :
\begin{itemize}
\item The matrix $R$ is a random orthogonal matrix.
\item The matrix $R$ is given by eq ( \ref{FOU} ).
\end{itemize}

The second case is a particular instance of the first one, exactly in the same way that a sequence 
of all zeros is a particular instance of a random sequence.

The first model can be studied using the replica method and one finds results very similar to those 
of the $p$-spin model we have already studied.

Now it can be proven that the statistical properties of the second model are identical to those of 
the first model, with however an extra phase.  In the second model (at least for some peculiar value 
of $N$, e.g.  $N$ prime \cite{MPR,BGU} ) there are configurations which have exactly zero 
energy.  These configuration form isolated valleys which are separated from the others, but have 
much smaller energy and they have a very regular structure (like a crystal).  An example of these 
configurations is
\be \si_k \equiv_{ N }k^{(N-1)/2} \ee 
(The property $k^{(N-1)}\equiv 1$ for prime \bd N \ed , implies that in the previous equations 
$\si_k=\pm 1$).  Although the sequence $\si_k$ given by the previous equation is apparently random, 
it satisfies so many identities that it must be considered as an extremely ordered sequence (like a 
crystal).  One finds out that from the thermodynamical point of view it is convenient to the system 
to jump to one of these ordered configurations at low temperature.  More precisely there is a first 
order transition (like a real crystalization transition) at a temperature, which is higher that the 
dynamical one.

If the crystallisation transition is avoided by one of the usual methods, (i.e.  explicit 
interdiction of this region of phase space or sufficient fast cooling), the properties of the second 
model are exactly the same of those of the first model.  Similar considerations are also valid for 
other spin models \cite{Frhe,MPRII,MPRIII} or for models of interacting particles in very large 
dimensions, where the effective range of the force goes to infinity \cite{CKPR,CKMP} .

We have seen that when we remove the quenched disorder in the Hamiltonian we find  a quite positive 
effect: a crystallisation transition appears like in some real systems.  If we neglect 
crystalization, which is absent for some values of $N$, no new feature is present in system without 
quenched disorder.  

These results are obtained for long range systems. As we shall see later the equivalence of short 
range systems with and without quenched disorder is an interesting and quite open problem.

\section{Some considerations on real glasses}

Here we will select some of the many characteristics of glasses we think are important and should be 
understood.  The main experimental findings about glasses that we would like to explain are the 
following:
\begin{itemize}
\item If we cool the system below some temperature ($T_G$), its energy depends on the cooling rate 
in a significant way.  We can visualize $T_G$ as the temperature at which the
relaxation times become of the order of a hour.

\item No thermodynamic anomaly is observed: the entropy (extrapolated at ultraslow cooling) is a 
linear function of the temperature in the region where such an extrapolation is possible.  For 
finite value of the cooling rate the specific heat is nearly discontinuous.  Data are consistent 
with the possibility that the true equilibrium value of the specific heat is also discontinuous at a 
temperature $T_c$ lower than $T_G$.  The difference of the entropy among the glassy phase and the 
crystal seems to vanish approximately at $T_{c}$.

\item The relaxation time (and quantities related to it, e.g.  the viscosity) diverges at low 
temperature.  In many glasses (the fragile ones) the experimental data can be fitted as 
\begin{eqnarray} 
\tau =\tau_0 \exp(\beta B(T))\\ 
B(T) \propto (T-T_c)^{-\la} 
\end{eqnarray}
where $\tau_0 \approx 10^{-13} s$ is a typical microscopic time, $T_c$ is near to the value at which 
we could guess the presence of a discontinuity in the specific heat and the exponent $\la$ is of 
order 1.  The so called Vogel-Fulcher law \cite{VF} states that $\la=1$.  The precise value of $\la$ 
is not well determine.  The value $1$ is well consistent with the experimental data, but different 
values are not excluded.
\end{itemize}

Now the theoretical interpretation of these results is quite clear (quite similar phenomena happens 
also in spin glass models in the mean field limits, where they can have been very carefully studied 
both numerically and analytically).  When we go near to the glassy phase the system may be frozen in 
many different configurations whose number is exponentially large.  The entropy density can thus be 
written as

\be S(T)=S_{c}(T)+W(T), \ee 
where $S_{c}(T)$ is the entropy inside each of these configurations (which is likely not too 
different from that of the crystal) and $W(T)$ is proportional to the logarithm of the total number 
of configurations (i.e.  it is the configurational entropy, called sometimes also complexity
\cite{THEO} ).  The
complexity vanishes linearly when $T$ approaches $T_c$ and remains zero at smaller temperature.  
This fact implies a the presence of thermodynamic transition at $T=T_{c}$.

The viscosity 
is dominated by the hopping from one equilibrium to an other equilibrium 
configuration.  The number of equilibrium configuration decrease when $T \to 
T_{c}$ and therefore the number of particles that are involved in each hopping 
process must increase.  The free energy barriers for such a process increases at 
the same time and diverges at $T_{c}$; the precise way in which they diverge it 
is a highly debated problem.

A consequence of this picture is that below $T_{c}$ there number of different equilibrium 
configurations (which contribute to the partition function) should not be anymore exponentially 
large.  In this region replica symmetry should be broken, i.e.  an arbitrarily small force should be 
enough to keep together two different replicas for an arbitrary large time.

The replica formalism seem therefore well suited to capture the phase transition in glasses.  At 
this moment it is not clear which is best way in which this formalism must be used.  We shall see in 
the next section a possible application.

\section{Toward real glasses}

It is interesting to find out if one can construct approximation directly for the glass transitions 
in liquids.  Some progresses have been done in the framework of mean field theory in the infinite 
dimensional cases.  Indeed the model for hard spheres moving on a sphere can be solved exactly in 
the high temperature phase when the dimension of the space goes to infinity in a suitable way
\cite {CKPR,CKMP}.

One of the most interesting results is the suggestion that the replica method can be directly 
applied to real glasses.  The idea is quite simple.  We assume that in the glassy phase a finite 
large system may state in different valleys, labeled by $\gamma$.  The probability distribution of 
the free energy of the valley is given by eq.  (\ref{FREE}).  We can speak of a probability 
distribution because the shape of the valleys and their free energies depends on the total number of 
particles.  Each valley may be characterized by the density
\be
\rho(x)_{\gamma}\equiv< \rho(x)>_{\gamma}
\ee
In this case we can define two correlation functions 
\bea
g(x)=\frac{\int dy < \rho(y) \rho (y+x)>_{\gamma}}{V}\\
 g_{R}(x)=\frac{\int dy < \rho(y)>_{\gamma} < \rho (y+x)>_{\al}}{V}, \label{gdef}
\eea
 where for simplicity we have assumed that the density of the particles is 1.  
 
A correct description of the low temperature phase must take into account both correlation 
functions.  The replica method does it quite nicely: $g$ is the correlation function inside one 
replica and $g_{R}$ is the correlation function among two different replicas.
\bea
g(x)=\frac{\int dy < \rho_{a}(y) \rho _{a} (y+x>}{V}\\
g_{R}(x)=\frac{\int dy < \rho_{a}(x) \rho _{b} (y+x)>}{V} \ \ mbox{with}\ \ a \ne b,
\eea
where here $a$ and $b$ label the replica and $\rho_{a}(x)$ is the density of particles of the 
replica $a$.

This would be more or less the traditional approach if we assume the existence of only one valley 
for the system.  On the contrary if we assume that the state may be in many different valleys we 
have to say something about the free energy distribution of free energy of the valleys.  This can 
be done using the replica method and using the variable $m$, defined in eq.  (\ref{DEFM}), which 
characterize the probability distribution of the free energy of those valley having a value of the 
free energy near the minimum.

 The problem is now to write closed equations for the two correlation functions $g_{R}$ and $g$.  
Obviously in the high temperature phase we must have that $g_{R}=1$ and the non-vanishing of 
$g_{R}-1$ is a signal of entering in the glassy phase.

The first attempt in this direction was only a partial success \cite{MEPA} and it will be described 
in the appendix.  A generalized hypernetted chain approximation was developed for the two functions 
\bd g \ed and $g_{R}$.

The replica formalism provides and automatic bookkeeping of all complications which would arise from 
the existence of many states.  If one assumes a given form for replica symmetry breaking, it 
correspond to a given form for the probability distribution of the $w$.  In the simplest case, 
called one step breaking, one divides the $n$ replicas in $n/m$ groups of $m$ replicas and one 
assumes the following structure of correlation functions:
\bea
<\rho_a(x) \rho_a(y)> &=& g(x-y) \ , \nn <\rho_a(x) \rho_b(y)> &=& g_{R}(x-y) \mbox{ for $a \ne b$ 
in the same group} \ ,\nn <\rho_a(x) \rho_b(y)> &=& 1 \mbox{ for $a$ and $b$ in different groups.}\ 
.
\label{ONESTEP}
\eea
Fortunately enough the results for the free energy density (per replica) do not depend on $n$ so 
that we do not have to specify this (unphysical)   parameter.

 A non trivial solution was found at sufficient low temperature both for soft and hard 
spheres and the transition temperature to a glassy state was not very far from the numerically 
observed one.  Unfortunately the value of the specific heat at low temperature is not the correct 
one (it strongly increases by decreasing the temperature).  Therefore the low temperature behaviour 
is not the correct one; this should be not a surprise because an esplicite computation show that the 
corrections to this hypernetted chain approximation diverge at low temperature.

These results show the feasibility of a replica computation for real glasses, however they point in 
the direction that one must use something different from a replicated version of the hypernetted 
chain approximation.  At the present moment it is not clear which approximation is the correct one, 
but I feel confident that a more reasonable one will be found in the near future.

\section*{Appendix}
We consider a system of $N$ interacting particle in a volume $V$.  We study the infinite volume 
limit in which $N \to \infty$ at fixed $\rho \equiv N/V$.  The Hamiltonian is given by:
\begin{equation}
	H(x)=\sum_{i \ne k} U(x_i-x_k).
	\label{Hamiltonian}
\end{equation}

We will consider a soft sphere case $U(x) =r^{-12}$.  We will work at density one and we will 
introduce the parameter $\Gamma\equiv \beta^{4}$.

 In this case the glass phase may be reached only with a 
very fast cooling rate, otherwise the system goes into the crystal phase.  This difficulty may be 
easily removed by considering a binary mixture, but this problem is not relevant here.

The hypernetted chain (HNC) approximation consists in considering only a given class of diagrams in 
the virial expansion \cite{HanMc}.  It gives a reasonable account of the liquid phase.  We will 
consider here this approximation because it has the advantage of having a simple variational 
formulation.  In the liquid phase, where the density is constant, the usual HNC equation (for the 
non replicated system) can be written as
\be
g(x) = \exp \left(-\beta U(x) + W(x)\right),
\ee
where:
\bea
g(x) &=& (1+h(x))= <\rho(x)\rho(0)>- \de(x),\\
W(x)&\equiv& 
 \int {d^dp \over (2 \pi)^d} \ e^{-ipx} 
{ {\bf h}(p)^2 \over 1+ {\bf h}(p)}\ , 
\label{HNC}
\eea
and we denote by ${\bf h}(p)$ the Fourier transform of $h(x)\equiv g(x)-1$.

This equation can be derived by minimizing with respect to $g(x)$ the following free energy per unit 
volume, in the space of functions of $|x|$:
\be
\beta F = \int d^dx \  g(x)[\ln (g(x))-1 +\beta U(x)] +
 \int {d^dq \over (2 \pi)^d} L({\bf h}(q)),
\ee
where $L(x) \equiv -\ln(1+x)+x-x^2/2$.

\begin{figure}
  \epsfxsize=400pt\epsffile[22 206 565 566]{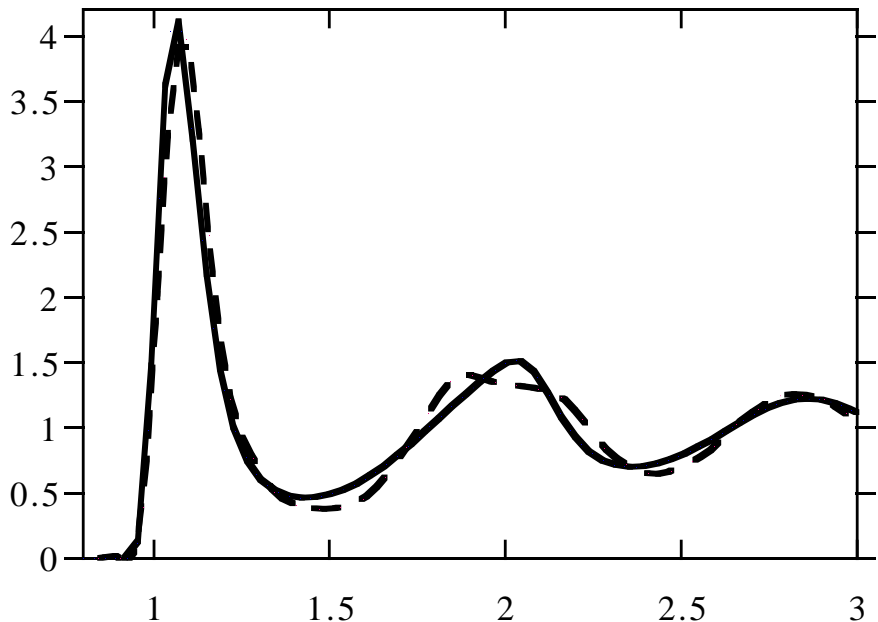} \caption[a]{\protect\label{DUE} The correlation 
  function of a system of soft spheres as a function of the distance at the dimensionless inverse 
  density $\Gamma =1.6$ corresponding to the glassy transition: numerical simulations (points) and 
  replica symmetric HNC equation (full line).  }
\end{figure}

The HNC equation gives a description of the liquid phase which is not perfect, but precise enough 
for our purpose: The energy (or equivalently the pressure), does not depart more than 15\% from the 
correct value, and the correlation function is also well
reproduced (see fig. ( \ref{DUE} )).

In \cite{MEPA} it was proposed a bold generalization of the HNC equations for $n$ replicas.  The 
replicated free energy is now
\be
\beta F =  \int d^dx \sum_{a,b} g_{ab}(x)
[\ln (g_{ab}(x))-1 +\beta  U(x) \delta_{a,b} ] +\mbox{Tr} L ({\bf h}),
\ee
where ${\bf h}$ is now an operator both in $x$ space and in replica space.

If one finds that at low enough temperatures there is a solution of the variational equations $\de F 
/\de g_{a,b}=0$, where $g_{ab}$ is non zero off the diagonal, replica symmetry is broken.  In the 
case where $g_{ab}$ is of the form shown in eq.  (\ref{ONESTEP}), this equation can be used to 
compute the properties of the correlation function in the glassy phase.  This approach amounts to a 
study of the density modulations in the glass phase at the level of the two point function.  In the 
glass phase $\rho_\ga(x)$ becomes space dependent.  However, as argued in the introduction, the 
necessity of averaging over the states $\ga$ forces us to study this $x$ dependence at the level of 
the two point correlations.  So our correlation $g_{R}$ reflects the structure of $\rho_\ga(x)$ as a 
sum of peaks of unit weights, smoothed by the average over states.

Within the one step breaking scheme (\ref{ONESTEP}), the free energy per replica is:
\bea
\beta F =  \int d^d x \{g(x)[\ln(g(x))-1 +\beta U(x) ] -(1-m) g_R(x)[\ln(g_R(x))-1] \}
\\
-\int {d^d q \over (2\pi)^d} \  \((
\frac12 {\bf h}(q)^2-\frac12 (1-m){\bf h}_R(q)^2 -{\bf h}(q)  \right.
\nn
\left.
+{1 \over m } \ln[1+ {\bf h}(q)-(1-m) {\bf h}_R(q)]-{1-m \over m } \ln[1+ {\bf h}(q)- {\bf h}_R(q)].
\))
\eea
Two transitions can be found: the static and  the dynamical transition.

\begin{itemize}
\item
The static transition is identified as the temperature (or density) at which there exists a non
trivial solution to the replicated HNC equations :
\be {\de F \over \de g(x)=0} \, \ \ \ {\de F/ \de g_R(x)=0} \label{Wdef}\ee
and 
\be \partial F/ \partial m=0, \ee 
for $ m \in [0,1]$ (in fact we must minimize the free energy with respect to $g$, but maximize 
with respect to $g_R$ and $m$).
\item
The dynamical transition may be characterized as the highest temperature at which there is a non 
trivial solution of the two stationarity equations $ \de F/ \de g(x)=0$ and $ \de F/
\de g_R(x)=0$ at $m=1^-$.  The corresponding equations are obtained by substituting $m \to 1$
in the first two equations of (\ref{Wdef}).  The equation for $g$ is identical to the usual HNC 
equation (\ref{HNC}), while $g_R$ is a solution of $g_R(x) = \exp(W_R(x))$, with $W_R$ can be 
extracted from the second equation of (\ref{Wdef}) at $m=1$.
\end{itemize}

\begin{figure}
  \epsfxsize=400pt\epsffile[22 206 565 566]{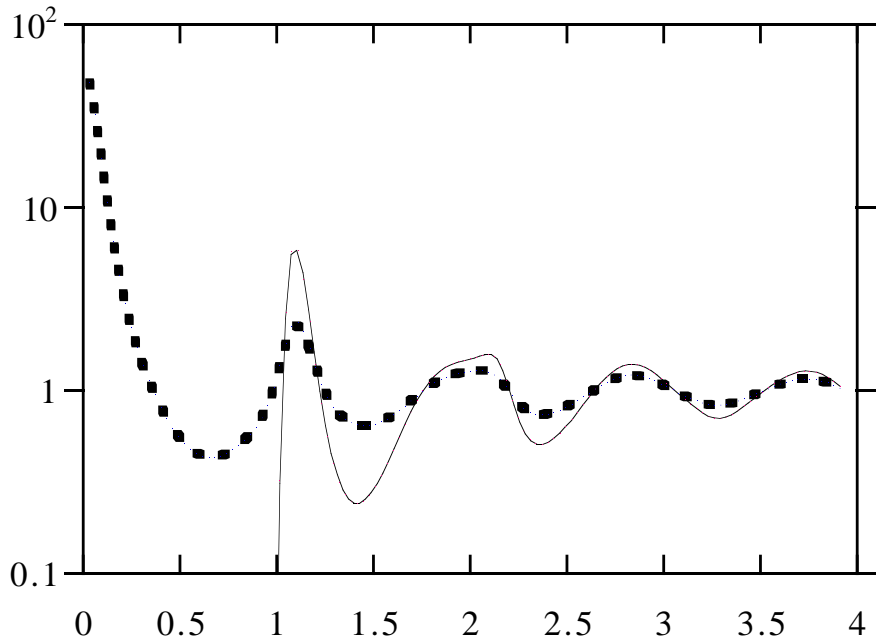} \caption[a]{\protect\label{TRE} The correlations 
  $g(r)$ (full line) and $g_{R}(r)$ (broken line) as functions of the distance, for soft spheres at the 
  density where replica symmetry is broken (e.g.  $\gamma=2.15$) }
\end{figure}

If we look for the numerical solution of the replicated HNC equations, we find a dynamical 
transition at $\Gamma\simeq 2.05$, and a static replica symmetry breaking solution at $\Gamma \simeq 
2.15$.  In numerical simulations the glass transition is found at a smaller value of $\Gamma$, 
namely $\Gamma=1.6$.  In the glass phase, the correlation function $g_1(r)$ is essentially a 
smoothed form of the function $g(r)$ plus an extra contribution at short distance which has integral 
near to $1$ (see Fig.2).  This form seems very reasonable: considering the definition (\ref{gdef}) 
of $g_R$, we see that it basically characterizes the average over $\ga$ of the product 
$\rho_\ga(x)\rho_\ga(y)$, which is precisely expected to have this kind of peak structure.

In spite of this nice form for $g_R$, this solution has some problems.  A first one is found on the 
value of the energy.  Although there is a discontinuity in the specific heat at $T_{c}$, it is 
extremely small and the final effects on the internal energy are more or less invisible.  The 
specific heat remains extremely large.  Moreover the value of $m$ has a very unusual dependence on 
the temperature.  In all the known models with one step replica symmetry breaking, the breakpoint 
$m$ varies linearly with $T$ at low temperatures.  Here we have a very different behaviour.  We have 
also computed the dynamical internal energy and found out that it differs from the equilibrium one 
by an extremely small account.
 
If we consider the qualitative behavior of the correlation functions, we find a reasonable form, on 
the other hand the energy in the glassy phase turns out to be quite wrong.  This computation should 
be considered only as a first step and work is in progress to find another approximation (different 
from the replicated HCN) which should be similar in spirit, but technically more sound.

\end{document}